\documentclass{eptcs}
\usepackage{pdfpages}
\usepackage{verbatim}
\usepackage{combelow}
\usepackage[show]{ed}
\usepackage{graphicx}

\newtheorem{theorem}{Theorem}

\newtheorem{definition}{Definition}

\newtheorem{example}{Example}

\begin{document}
\author{J.H. Davenport\\Department of Computer Science\\University of Bath, Bath BA2 7AY, U.K.\\\tt masjhd@bath.ac.uk}
\title{Formal Methods and CyberSecurity}
\date{21 June 2019}\def\authorrunning{J.H. Davenport}
\def\titlerunning{Formal Methods and CyberSecurity}
\maketitle
\begin{abstract}
Formal methods have been largely thought of in the context of safety-critical systems, where they have achieved major acceptance. Tens of millions of people trust their lives every day to such systems, based on formal proofs rather than ``we haven't found a bug'' (yet!). Why is ``we haven't found a bug'' an acceptable basis for systems trusted with hundreds of millions of people's personal data?

This paper looks at some of the issues in CyberSecurity, and the extent to which formal methods, ranging from ``fully verified'' to better tool support, could help. Alas \cite{RoyalSociety2016a} only recommended formal methods in the limited context of ``safety critical applications'': we suggest this is too limited.
\end{abstract}
% cybersecurity; input sanitisation; taint analysis

\section{Introduction}
CyberSecurity\footnote{The precise definition of CyberSecurity is debatable: we can take is as failures of security, generally defined as ``preserving the CIA --- Confidentiality, Integrity and Availability'' of digital information, where computer system played a critical part in the failure.} failures abound, and the number of people that can be affected by even a single failure is amazing --- 148 million for Equifax \cite{Bloomberg2018b} and probably more for the Starwood\footnote{Generally called ``Marriott'', but in fact due to the Starwood chain before Marriott took it over.} breach: \cite{BBC2018o} states 500 million, but \cite{Irwin2019b} ``downgrades'' this to 383 million.  The financial costs can be substantial: bankruptcy in the case of American Medical Collection Agency \cite{Ford2019b} and a provisional \pounds183M fine for British Airways \cite{Guardian2019i}.  These problems have attratced attention at the highest scientific levels \cite{RoyalSociety2016a}.
\par
There are many reasons for CyberSecurity failures, and even a given failure may have multiple causes. For example, the U.S.~Government investigation \cite{GAO2018a} into Equifax states ``Equifax's investigation of the breach identified four major factors
including identification, detection, segmenting of access to databases, and data
governance that allowed the attacker \dots''.  However, none of these would have been triggered had it not been for the original bug in the Apache code \cite{Lenart2017a}, which was of the well-known (Number 1 Application Security Risk in \cite{OWASP2017a})  family of ``Injection'' (or ``Remote Code Execution'') attacks, and which would probably have been detected by an automatic taint analysis tool such as \cite{LivshitsLam2005}.
\par
Though attributing causes at scale is difficult, a well-known textbook \cite{McGraw2006} claims that about 50\% of security breaches are caused by coding errors. Hence it behoves security practitioners to look seriously at coding errors, while recognising that this is only one facet of the problem. This is taken up by the Payments Card Industry in \cite{PCI2018b}, essentially the only world-wide mandatory security standard, in two requirements.
\begin{description}
\item[6.5]Address common coding vulnerabilities in software-development processes as follows:
\begin{itemize}
\item Train developers at least annually in up-to-date secure coding techniques, including how to avoid common coding vulnerabilities;
\item Develop applications based on secure coding guidelines.
\end{itemize}
\item[6.6]For public-facing web applications, address new threats and vulnerabilities on an ongoing basis and ensure these applications are protected against known attacks by either of the following methods:
\begin{itemize}
\item Reviewing public-facing web applications via manual or automated application vulnerability security assessment tools or methods, at least annually and after any changes;
%Note: This assessment is not the same as the vulnerability scans performed for Requirement 11.2.
\item Installing an automated technical solution that detects and prevents web-based attacks (for example, a web-application firewall) in front of public-facing web applications, to continually check all traffic.
\end{itemize}
\end{description}
It is noteworthy that, despite apparently insisting on secure coding in 6.5, they require the additional defences in 6.6, realising that \emph{errare humanum est}, and the 6.5-developed code may not actually be secure.  Is it possible (the author thinks so, but the experiment has yet to be performed) that adding formal methods to 6.5 would render 6.6 %(at least the first option)
 redundant?
Full formal verification of a complete system should certainly suffice.
\begin{quote}
Complete formal verification is the only known way
to guarantee that a system is free of programming
errors. \cite[describing seL4: a verified operating system]{Kleinetal2009a}
\end{quote}
Such a verified operating system has been used in medical devices, but probably not sufficiently widely, as 500,000 already-fitted pacemakers have had to be upgraded through security weaknesses \cite{Guardian2017g}, and insulin pumps are also vulnerable \cite{Newman2019k}. See \cite{Heiser2019a} for a recent update on seL4.  However, most of us do not have the opportunity to start from scratch, and have to live on top of imperfect, unverified systems, interoperating with other systems via large, generally unverified, protocols, such as TLS. 
\section{TLS and its issues}
The TLS protocol (and its predecessor SSL) are the basis of most Internet security, underpinning, for example, \verb+https+. They also have displayed some of the most prominent problems.
\begin{description}
\item[Correctness]Paulson \cite{Paulson1999} ``Proved TLS Secure'', according to folklore. More precisely, the abstract states ``All the
obvious security goals can be proved'', but the paper itself is more more nuanced.
\begin{quote}
Is TLS really secure? My proofs suggest that it is, but one should draw no conclusions without reading the rest of this paper, which describes how the protocol
was modelled and what properties were proved. I have analyzed a much simplified
form of TLS; I assume hashing and encryption to be secure.
\end{quote}
There has been much work on TLS security since, e.g. \cite{Heetal2005,Krawczyketal2013a}. In particular, the latter used `real' encryption, the RSA PKCS \#1 v1.5, recommended, rather than `ideal' encryption. Again, these all focus on the idealised protocol, rather than implementations.
\item[Heartbleed \cite{NIST2014a}]This, arguably the most serious security issue of 2014, at least as perceived by the media (for example \cite{BBC2014b}) and the public (for example \cite{Steinberg2014a}), was a bug in a particular, but very widely used, implementation (OpenSSL) of TLS, and hence instantly falls outside the scope of \cite{Paulson1999}. Furthermore, it was a bug in an extension \cite{Seggelmannetal2012a} which postdates \cite{Paulson1999}.  \cite{Seggelmannetal2012a} states ``This document does not introduce any new security    considerations'', which is also true.
\par
The bug itself was a bounds checking bug, and thus could have been flagged by even relatively weak static analysis tools.  Looking a bit deeper, it was caused by assuming that the other end was behaving correctly. This seems to be a general class of errors, oddly missing from \cite{AndersonNeedham1995}.
\item[Poodle \cite{Molleretal2014a,NIST2014b}]This also appeared in 2014. It requires two ``features'' to operate.
\begin{enumerate}
\item Many TLS implementations contain ways to downgrade to SSL 3.0 if the other end doesn't support TLS itself. However this downgrade (again, a feature of implementations, so outside the scope of \cite{Paulson1999}) is typically not a proper protocol negotiation, and can be subverted by an active attacker. As \cite{AndersonNeedham1995} state ``where the identity of a principal is essential to the meaning of a message, it should be mentioned explicitly in the message'', and indeed should be authenticated.
\item Once downgraded to SSL 3.0, the attacker can exploit this.
\begin{quote}
The most severe problem of CBC encryption in SSL 3.0 is that its block cipher padding is
not deterministic, and not covered by the MAC (Message Authentication Code): thus, the
integrity of padding cannot be fully verified when decrypting. \cite{Molleretal2014a}
\end{quote}
Paulson \cite{Paulson1999} states, not unreasonably, ``I assume hashing and encryption to be secure'', as this is a separate set of proof technologies, and generally only produces relative security proofs.
\end{enumerate}
\end{description}
Hence we have a proof of correctness of a (simplified, but in fact the simplification is irrelevant here) version of an abstract protocol, and major bugs in implementations. One is a ``coding'' bug, while the other is a combination of a protocol bug and a cryptography bug.  Though not directly relevant to this paper, \cite{Salz2017a} demonstrates that Heartbleed (and Poodle) had a major positive effect on the OpenSSL project.
\section{Agile versus Secure}
``Agile Development'' \cite{Becketal2001} is a major theme in software development. 
Mark Zuckerberg can be said to have taken this theme to the extreme in 2009.
\begin{quote}
``Move fast and break things'' is Mark's prime directive to his developers and team. ``Unless you are breaking stuff,'' he says, ``you are not moving fast enough.''  \cite{Blodget2009a}
\end{quote}
In both safety-critical and security-conscious programming, ``breaking things'' comes with a very high price. Aeroplanes can't be uncrashed, and data can't be unleaked. 

The problems with using ``Agile'' methods in security are well-documented, at practitioner level, e.g. a recent ``Security + Agile = FAIL'' presentation \cite{Lane2018a}, in many theoretical analyses as well as the interview-based research in \cite{Bartsch2011a} for small teams and  \cite{vanderHeijdenetal2018a} for large multi-team projects. Both mention team expertise in security as a significant problem.
\begin{description}
\item[\cite{Bartsch2011a}]The
overall security in a project depends on the security expertise
of the individuals, either on the customer or developer side.
This corresponds to the agile value of ``individuals and
interaction over processes and tools'' \cite[Value 1]{Becketal2001}.
\item[\cite{vanderHeijdenetal2018a}]The interviewees
generally agree that more could be done to provide security education
and training to employees. Without prompting, several
interviewees mentioned training as an important factor for increasing
security awareness and expertise. 
\end{description}
It is very hard to take security seriously in this setting.
\begin{description}
\item[\cite{Bartsch2011a}]security ``is only of interest [to the customer] when money-aspects are
concerned''.
\item[\cite{vanderHeijdenetal2018a}]One Test Manager articulated his team view
that ``security is not currently seen as part of working software,
it only costs extra time and it doesn't provide functionality''.  With less focus on
providing extensive (security) documentation typical for agile, ineffective
knowledge sharing between security officers and agile team
members is especially problematic.
\item[\cite{TahaeiVaniea2019a}\footnote{A more general survey, but many papers surveyed were ``Agile''.}]``Security is often referred to as a NFR [non-functional requirement] in that it
is expected to be included as part of high quality code development,
but is rarely listed as an explicit requirement.
As a result, developers prioritise security below more-visible
functional requirements or even easy-to-measure activities
such as closing bug tracking tickets.''
\end{description}
\par\noindent
It would be tempting to conclude that ``Agile'' and ``Secure'' are, or at least are close to being, mutually contradictory. But there has been some analysis of the same apparent contradiction in the safety-critical industry \cite{Chapman2016a}.  Other than ``Embedded Systems''\footnote{Actually, Embedded Systems are a comparatively neglected, but important, CyberSecurity area. See, for example, \cite{OConnoretal2019a} for a description of a pervasive design fault in the ``home security'' market.} \cite[\S 3.6]{Chapman2016a}, this analysis of the problems is fairly close to the practitioner view in \cite{Lane2018a}, and we could reasonably ask what lessons could be carried across. 
\section{The Need for Tools}
\def\foo{\cite[\S 4.1]{Chapman2016a}}
\def\bar{\cite[\S 6]{Chapman2016a}}
There are two key points.
\begin{description}
\item[\foo]Strong static verification tools tend to complement (not replace) human-driven review\footnote{A point made in the context of XP and Agile in 2004 \cite{Wayrynenetal2004}.}. The tools are very good at some problems (e.g. global data flow analysis, theorem proving) where humans are hopeless, and vice versa. If we do the static verification first, then we can adjust manual review processes and check-lists to take advantage of this.
\item[\bar]The sixty-four-million-dollar-question, it seems, is how much ``up-front'' work is ``just right'' for a particular project. We doubt there’s a one-size-fits-all approach, but surely the answer should be informed by disciplined requirements engineering of non-functional properties (e.g. safety, security and others) that can inform the design of a suitable architecture and its accompanying satisfaction argument.
\end{description}
\par\noindent
Facebook grew, security (and ``product quality'' in general: it is not clear whether security was the main driver here) became more important, and by 2014 Zuckerberg had changed his views.
\begin{quote}
``Move fast with stable infrastructure.'' It ``may not be quite as catchy as `move fast and break things,''' Zuckerberg said with a smirk. ``But it's how we operate now.'' \cite{Statt2014a}
\end{quote}
\par\noindent
One might think his views were converging with the views of  \cite{Chapman2016a}. However, the Heartbleed story should remind us that the fact that a modification ``has no new security considerations''  \emph{as designed}  \cite{Seggelmannetal2012a} doesn't mean that an implementation of that idea has no new security considerations. Hence the call in \foo{} for strong static verification tools. Such tools are generally seen as expensive and slowing down the development process, but \cite{BrainSchanda2012a} shows that they need not be. In particular, they show that, for a real application (890,000 physical lines of Ada code), the cost of incremental verification can be reduced from ``nightly'' to ``coffee'', and hence can reasonably form part of a continuous integration toolchain, as is done at the company studied in \cite{BrainSchanda2012a}. Readers might  comment that their own applications are not in Ada, but \cite[\S5.6]{ChapmanMoy2018a} discusses mixed-language programming, especially with C.  A similar point is made in \cite{Distefanoetal2019a}, describing the Infer tool running on Java/Objective C/C++, where moving from overnight reporting to near real-time reporting moved the fix rate from 0\% to 70\%.
\par
That these techniques are reaching the mainstream of CyberSecurity can be seen from Amazon Web Services adoption of them  \cite{Vogels2019a}, Google \cite{Sadowskietal2018a}, Facebook \cite{Distefanoetal2019a}, and the recent DefectDojo release by OWASP \cite{OWASP2019b}.
\section{The Scope of Tools and Formal Methods}
There is a substantial range of tools, and degrees of formality, and \bar{} is probably correct in saying ``We doubt there’s a one-size-fits-all approach''. At one extreme, there are the humble, but still surprisingly effective, \verb+lint+ and its equivalents, looking, essentially, for dangerous or dubious, though legal, syntax. 
\subsection{Ada and SPARK}
At the other extreme, there are languages, such as the SPARK Ada subset \cite{ChapmanMoy2018a} designed with verification in mind and heavily employed in the safety-critical sector such as railways and air traffic control, which can also be deployed for demanding secure applications, such as an RFC4108-compliant \cite{Housley2005} secure download system for embedded systems \cite{Chapman2018b}.
\subsection{C/C++}
There is, however, a large middle ground between these two extremes.  Even if the application is required to be in C or C++, there is a lot to be said for sticking to a safer (even if not provably safe) subset of the language \emph{and associated libraries}, such as eschewing \verb+strcpy+ in favour of \verb+strncpy+. This can often be enforced by static verification tools.
We note that Google's ``Zero Day'' project reports \cite{Google2019g} that 68\% of all such zero-day exploits (i.e. exploits discovered in the wild first) were caused by memory corruption errors, and Microsoft report a very similar story \cite{Thomas2019a}.

There is a good survey of such subsets and standards in \cite[Appendix F]{CPNI2019a}.   As that notes, the ISO standard for secure C coding \cite{ISO2013d} has the unusual (for this middle ground) but important concept of ``taint analysis'' (as in \cite{LivshitsLam2005}): input data should be considered ``tainted'' until it has been sanitised.  This is particularly important for network-oriented applications, where it is natural for the programmer to believe that the other party is behaving correctly (see {\bf Heartbleed} above).
\subsection{Java}
Closer to the SPARK Ada end of the spectrum we find Safety-Critical Java \cite{Cavalcantietal2017a}.  The author does not have enough experience with this to comment directly. However, the Java ecosystem (Stack Overflow etc.) is far from security-aware \cite{Mengetal2018a}.                                                                                                                                                                                                                                                                                                                                                   The fact that an application is in Java doesn't mean it's free from security coding errors: see \cite{Google2018a} for a recent example.
\par
There is a static analysis security tool for Java described in \cite{LivshitsLam2005}.
As with \cite{ISO2013d}, this has ``taint analysis'' as its major feature, and at the time it spotted some significant-seeming problems.

\subsection{JavaScript}
JavaScript is a particular problem for Security. There are some verification tools, e.g. GATEKEEPER as described in \cite{GuarnieriLivshits2009}. However, even if it were possible to guarantee a particular piece of stand-alone JavaScript, that is not how the current paradigm operates. As \cite{MeyerovichLivshits2010a} writes:
\begin{quote}
Much of the power of modern Web comes from the
ability of a Web page to combine content and JavaScript code
from disparate servers on the same page. While the ability
to create such mash-ups is attractive for both the user and
the developer because of extra functionality, code inclusion
effectively opens the hosting site up for attacks and poor
programming practices within every JavaScript library or API
it chooses to use.
\end{quote}
Though not explicit in this statement, an additional weakness is that this combination is \emph{dynamic}.
%: a property exploited by the recent Magecart attack on British Airways \cite{Barth2018a,Guardian2019i}, where an external library the British Airways code incorporated was hacked.    %JHD not quite, the library was loaded from BA's copy.
The obvious solution would be some kind of sandboxing of the external resources relied upon, but the nature of JavaScript makes this difficult.  \cite{Maffeisetal2009} describe one such sandboxing, but it only works for a subset of JavaScript and relies on a combination of filtering, rewriting and wrapping to guarantee security. That it can do so at all is a remarkable feat of formal methods, given that previous attempts such as Facebook's FBJS have subtle flaws \cite{MaffeisTaly2009}, and that the formal semantics of JavaScript being relied upon are very much a piece of reverse engineering.
\par
In fact the dynamic loading from multiple sites is often not good for performance, and web performance engineers recommend tools to bundle the pages: this could usefully be combined with the sort of protection described by \cite{Maffeisetal2009}.
\par
An alternative solution is used by Google,who are introducing a form of taint analysis into Chrome \cite{Kotowicz2019a} through run-time typing. When enabled, this means that the 60+ dangerous DOM API functions can only be called with arguments whose type is that emitted by \verb+TrustedTypes+ functions. Google expects that these functions would be manually verified, but this does open the door to formal verification of \emph{certain} security policies in what is currently a very challenging environment for formal methods. We note the complex interaction between
\begin{itemize}
\item the  server
\end{itemize}
\section{Education}
\cite[Requirement 6.5]{PCI2018b} called for education of developers. Education of mainstream programmers, as opposed to CyberSecurity specialists, in CyberSecurity has been neglected until recently, and this neglect has been lamented as far as the Harvard Business Review \cite{Cable2019a}. Developments in professional accreditation are changing this \cite{Cricketal2019a}.  However, there are limitations, even beyond  \emph{errare humanum est}, in relying on education.
\begin{enumerate}
% \yen : yen mark (taken from ascmac.sty for jlatex)
\def\yen{{\setbox0=\hbox{Y}Y\kern-.97\wd0\vbox{\hrule height.1ex
width.98\wd0\kern.33ex\hrule height.1ex width.98\wd0\kern.45ex}}}%
\item There is experimental evidence that both trained students \cite{Naiakshinaetal2018a} and professional developers \cite{Naiakshinaetal2019a} will ignore security considerations unless \emph{explicitly} instructed to take them into account. Lest this be thought to be a purely academic exercise with little relevance to the real world, consider the recent \yen55M password problem described in \cite{Cimpanu2019s}.
\item There is field evidence that explicit requirements such as \cite{PCI2018b} are ignored in practice, e.g. the Forever 21 breach \cite{Biscoe2018a}, or Macy's \cite{Blackmon2018a}. They may also not be communicated down the software supply chain, as in the Ticketmaster case \cite{Inbenta2018a}.
\item Many educational resources, both formal textbooks \cite{TaylorSakharkar2019a} and informal resources such as Stack Overflow \cite{Fischeretal2017a}, pay very little attention to security, and indeed can be positively harmful. The discussion in Stack Overflow (analysed in \cite[\S4.3.1]{Mengetal2018a}) of cross-site request forgery (CSRF --- this was in the OWASP top 10 in 2013 \cite{OWASP2013a}, but dropped from \cite{OWASP2017a} ``as many frameworks include CSRF defenses'') is especially worrying.  By default, Spring implicitly enables protection against this. But all the accepted answers to CSRF-related failures simply suggested disabling the check. There were no negative comments about this, and indeed a typical response is ``Adding \verb!csrf().disable()!
solved the issue!!! I have no idea why it was enabled by default''. 
\end{enumerate}
As we have noted, \cite{PCI2018b} both mandates education and does not rely solely on it.
\par
However, as the safety-critical community laments (at least in the U.K. and U.S.A.: cultures do differ here), there is very little training in formal methods for most undergraduates. 
\section{Conclusions}
As the media never tire of saying, there are far too many security breaches, and, though they have multiple causes, \cite{McGraw2006} claims that about 50\% of security breaches are caused by coding errors. There appears to be a culture of accepting these, with the U.S.~Government investigation \cite{GAO2018a} into Equifax blaming many factors but not the actual bug, and \cite{PCI2018b}  taking a ``necessary but not sufficient'' approach to education in secure coding.
\begin{description}
\item[Education]Could certainly do better \cite{Cable2019a}, though there are encouraging signs \cite{Cricketal2019a} and useful ideas when it comes to improving informal resources  \cite{Fischeretal2019a}. However, informal resources can be dangerous when it comes to security, and \cite{Cricketal2019a} recommends giving \emph{all} students the advice in \cite{Chenetal2019a}: ``If you pick up a SSL/TLS answer from Stack Overflow, there's a 70\% chance it's insecure''.

More training in formal methods would be welcomed, at least in those cultures where it is lacking.
\item[Customers/Managers]need to be much more upfront about security requirements \cite{Naiakshinaetal2018a,Naiakshinaetal2019a}, and enforce (e.g. by requiring tool support during any CI/CD process, such as \cite{BrainSchanda2012a} describe) at least ``middle ground'' requirements. In the case of outsourced development, explicit penalty clauses for failing penetration tests should concentrate the developers' minds.
\item[C/C++ people]These programmers should be much more aware of techniques for secure coding, such as those described in  \cite[Appendix F]{CPNI2019a}, and the various tools for static analysis.
\item[Java people]In view of the significance of injection attacks (Number 1 in \cite{OWASP2017a}), programmers should be aware of taint analysis, as in \cite{LivshitsLam2005}.
\item[JavaScript people]There are some techniques, such as \cite{Maffeisetal2009},  for protecting JavaScript applications, but they are not deployable in the the typical JavaScript ``dynamic loading web page'' environment. Furthermore this environment is basically antithetical to security, as British Airways is learning to the cost of \pounds183M \cite{Guardian2019i}.
\item[1)]Hence the first real challenge of JavaScript lies with the tool makers: there are, as far as the author knows, no JavaScript verifiers in existence, and no page-bundler that checks for version drift, or does incremental verification (which might be comparatively cheap, as in \cite{BrainSchanda2012a}).
\item[2)]An alternative approach might be to change the JavaScript model. This is advocated in \cite{Zhangetal2019c}, based on their analysis of what third-party scripts do in the wild.  This is not a completely radical idea: Google is testing its \verb+TrustedTypes+ feature \cite{Kotowicz2019a}, with the motivation ``The DOM API is insecure by default and requires special treatment to prevent XSS''.
\item[Empirical Research]There is not much analysis of the efficacy of various techniques in security programming. \cite{AustinWilliams2011a} compares various techniques, and states the following.
\begin{quote}
Based on our case study [of two large programs], the most efficient vulnerability discovery technique is automated penetration testing.  Static analysis finds more vulnerabilities but the time it takes to classify false positives makes it less efficient than automated testing.
\end{quote}
This assumes that ``false positives'' are acceptable, a debatable point of view.  It would be good to have more such research.
\item[Tool developers]There is a lack of tools (or at least a lack of awareness of tools) that can be neatly integrated into a security programming toolchain the way such tools are integrated in safety-critical toolchains \cite{BrainSchanda2012a}.
\end{description}
\par\noindent{\bf Acknowledgements:} The author is grateful to the Fulbright Programme for a CyberSecurity Scholarship, and to many correspondents and discussions, notably with Tom Crick, Alastair Irons and Tom Prickett; also Tim French. The FROM2019 referees made useful comments.
\bibliography{../../../jhd}
\end{document}